\documentclass[conference]{IEEEtran}
\IEEEoverridecommandlockouts
\usepackage{cite}
\usepackage{amsmath,amssymb,amsfonts}
\usepackage{algorithmic}
\usepackage{graphicx}
\usepackage{textcomp}
\usepackage{xcolor}
\def\BibTeX{{\rm B\kern-.05em{\sc i\kern-.025em b}\kern-.08em
    T\kern-.1667em\lower.7ex\hbox{E}\kern-.125emX}}
\begin{document}

\title{Learning with Knowledge of Structure: A Neural Network-Based Approach for MIMO-OFDM Detection
}

\author{Zhou Zhou, Shashank Jere, Lizhong Zheng, and Lingjia Liu
\thanks{Z. Zhou, S. Jere, and L. Liu are with ECE Department at Virginia Tech. Their work is supported in part by National Science Foundation (NSF) under grants CCF-1937487 \& CNS-2003059 and Intel Corporation. L. Zheng is with EECS Department at Massachusetts Institute of Technology. He is supported in part by the Office of Naval Research (ONR) under MURI Grant N00014-19-1-2621, NSF under grant CNS-2002908, and Intel Corporation. The corresponding author is L. Liu (ljliu@vt.edu).}
}


\maketitle

\begin{abstract}
In this paper, we explore neural network-based strategies for performing symbol detection in a MIMO-OFDM system. 
Building on a reservoir computing (RC)-based approach towards symbol detection, we introduce a symmetric and decomposed binary decision neural network to take advantage of the structure knowledge inherent in the MIMO-OFDM system. 
To be specific, the binary decision neural network is added in the frequency domain utilizing the knowledge of the constellation.
We show that the introduced symmetric neural network can decompose the original $M$-ary detection problem into a series of binary classification tasks , thus significantly reducing the neural network detector complexity while offering good generalization performance with limited training overhead. Numerical evaluations demonstrate that the introduced hybrid RC-binary decision detection framework performs close to maximum likelihood model-based symbol detection methods in terms of symbol error rate in the low SNR regime with imperfect channel state information (CSI).
\end{abstract}


\section{Introduction}
MIMO-OFDM is the primary air interface in modern communication standards such as 4G (LTE-Advanced) and 5G New Radio (NR)~\cite{parkvall2017nr}. 
As opposed to single-input single-output systems (SISO), 
MIMO-OFDM systems have symbols interfering in space and time due to the concurrent reception of signals at the receive antennas and the multipath propagation feature of wireless channels. 
In order to achieve the transmission (capacity) gains that MIMO offers, the choice of the symbol detection strategy plays a critical role. 
However, the following practical issues are present in conventional model-based symbol detection strategies: (1) Model Uncertainty - The exact characterization of the true end-to-end link from the transmitter to the receiver cannot be obtained. Instead, only a linear and static approximation of the wireless channel and a coarse approximation to the hardware imperfections in the transceiver are available. For e.g., Non-linearities in radio frequency (RF) circuits such as the power amplifier (PA) at the transmitter and finite quantization in baseband circuits such as analog-to-digital converter (ADC) at the receiver are hard to characterize accurately.
Additionally, channel state information (CSI) is not directly accessible and needs to be estimated at the receiver using the received signal. Algorithms can thus only use the estimated CSI, which yields diminishing detection success rate as the accuracy of the estimated CSI decreases, particularly in the low signal-to-noise ratio (SNR) regime~\cite{andrews2003optimum}. (2) Implementation Complexity - The computational complexity of optimal detection algorithms have an exponential relationship with the number of variables to be detected. Therefore, applications with large number of OFDM sub-carriers, high modulation orders and large number of transmit/receive antennas render the problem non-deterministic polynomial-time hard (NP-hard), and thus unable to meet latency requirements~\cite{papadimitriou1981complexity}.

The recent advent of neural networks (NNs) has led to the application of learning-based approaches to a wide array of scientific problems~\cite{lecun2015deep}.
Specifically, the concept of artificial intelligence (AI)-enabled cellular networks has been introduced~\cite{DBLP:journals/corr/abs-1907-07862}.
Such successful NN-based approaches motivate us to integrate NN strategies in symbol detection algorithms in order to mitigate the aforementioned issues from model-based approaches. 
However, this seemingly simple rationale is in fact extremely challenging due to the inherent constraints of wireless systems.
\begin{itemize}
\item \emph{Data Deficit}: For the purpose of training, we can only access a limited training dataset sent over-the-air in the physical layer (PHY). For instance, in 3GPP LTE/LTE-Advanced and 5G systems the demodulation reference signals in a $2 \times 2$ MIMO-OFDM system occupy approximately $10\%$ of the system resources in terms of resource elements (REs), whereas the remaining $90\%$ of REs are occupied by data symbols. Accordingly, the variance-bias tradeoff is insufficient to be guaranteed without an asymptotic large training dataset~\cite{yang2020rethinking}.
\item \emph{Learning Deficit}: Generic NNs are often slow to configure due to the large number of hyper-parameters that need to be optimized. Gradient descent-based learning can be slow to converge or may converge towards a solution with poor generalization performance~\cite{hardt2016train}. 
\end{itemize}
To address the above challenges, we consider incorporating inductive bias, i.e., prior knowledge, into the NN architecture by leveraging the structural properties of the MIMO-OFDM signal. This helps the network make inferences that go beyond the observed data, thus potentially solving the fundamental data deficit issue. The introduced NN structure leverages the following structural information of the MIMO-OFDM signal:
\begin{itemize}
\item Constellation of the digital modulation scheme.
\item Time-domain convolution and superposition operation by the wireless channel.
\item The presence of different frequencies inherent in the OFDM signal.
\end{itemize}
In the introduced NN, the constellation structure is reflected through a process which can generalize a binary detector to a multiple-class detector. 
The time-domain knowledge is captured through the dynamics of reservoir computing (RC), a special category of recurrent neural networks (RNNs)~\cite{RCMIMO1, zhou2019, zhou2020AAAI, zhou2020rcnet}. 
The frequency information is inherent in the weight initialization of the output layer of RC. Our evaluation results show that the introduced method performs close to maximum likelihood (ML) estimation in the low SNR regime while using imperfect (estimated) CSI.

The organization of this paper is as follows. In Sec.~\ref{ps}, we give an overview of the MIMO-OFDM symbol detection problem. In Sec. \ref{introduced_app}, we describe the technical details of our introduced method. Sec.~\ref{exp} evaluates the performance of our approach in comparison to conventional symbol detection methods as well as less other general neural network structures. Finally, we conclude the paper in Sec.~\ref{conclusion}.


\section{Problem Statement}
\label{ps}
\subsection{MIMO-OFDM System}
\begin{figure}
    \centering
    \includegraphics[height = 0.4\linewidth, width = 1\linewidth]{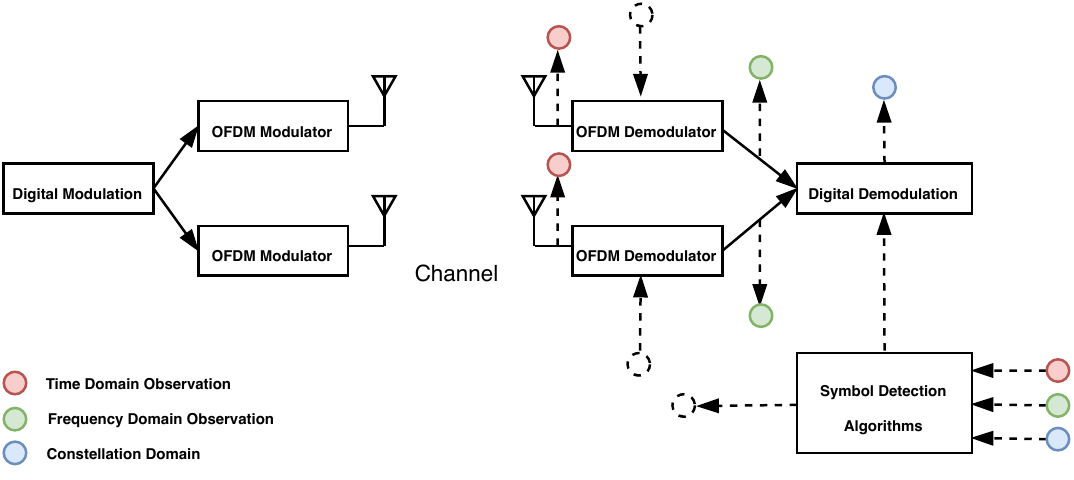}
    \caption{MIMO OFDM System}
    \label{fig_mimo_ofdm}
\end{figure}
Let $N_t$ and $N_r$ represent the number of transmit and receive antennas respectively; let $N_c$ represent the number of sub-carriers. 
Without loss of generality, we assume that the modulation scheme used is 16-QAM which is denoted by its alphabet as ${\mathcal A}= \{-3, -1, +1, +3\} \times \{-3j, -1j, +1j, +3j\}$. 
Let ${\boldsymbol X}\in {\mathcal A}^{N_t \times N_c}$ be the transmitted MIMO-OFDM symbol in the space-frequency domain, the received signal becomes
\begin{equation} 
\label{MIMO_OFDM0}
\boldsymbol{Y}=\sum_{\ell=0}^{L-1} \boldsymbol{H}[\ell] \boldsymbol{X} \boldsymbol{F}\boldsymbol{J}_{\ell} +{\boldsymbol N}
\end{equation}
where ${\boldsymbol F} \in {\mathbb C}^{N_c \times N_c}$ is the inverse discrete Fourier transform (IDFT) matrix, ${\boldsymbol{J}}_{\ell} \in {\mathbb C}^{N_c \times N_c}$ is a cyclic permutation matrix with $\ell$, $\boldsymbol N$ is the matrix of additive white Gaussian noise (AWGN), $L$ is the number of multipath components of the wireless channel. For ease of discussion, the Cyclic Prefix (CP) is removed in the model.

\subsection{Symbol Detection}
\begin{figure}[t]
\centering
\includegraphics[width =1 \linewidth]{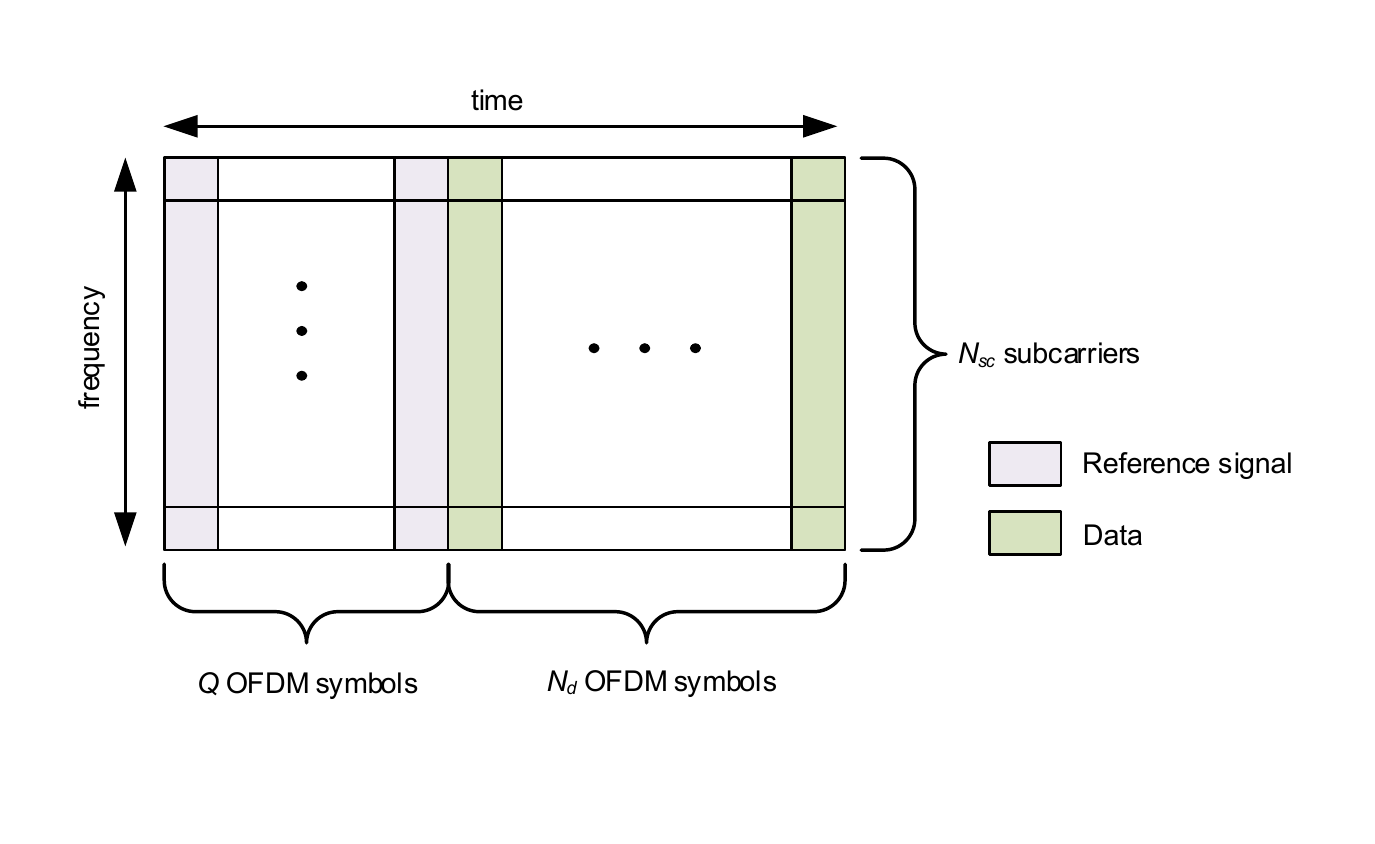}
\caption{Pilot and data structure}
\label{fig_training}
\end{figure}
The symbol detection task is to design an inverse map which infers $\boldsymbol X$ from the observation $\boldsymbol Y$. 
For simplicity, we consider using the real-valued version of $\boldsymbol Y$ and $\boldsymbol X$ in the inference task, which are respectively defined as 
\begin{align*}
{\boldsymbol {\tilde Y}} := \begin{bmatrix}
{\text{Re}}(\boldsymbol {Y})\\
{\text{Im}}(\boldsymbol {Y})
\end{bmatrix} {\text{and}}, 
{\boldsymbol {\tilde X}} := \begin{bmatrix}
{\text{Re}}(\boldsymbol {X})\\
{\text{Im}}(\boldsymbol {X})
\end{bmatrix}
\end{align*}
where $\boldsymbol {\tilde X}$ and $\boldsymbol {\tilde Y}$ now represent 4-PAM symbols, the definition of the real-valued equivalent receive signal model to (\ref{MIMO_OFDM0}) can be found in the Appendix. 
The pilot and data structure of an OFDM subframe is shown in Fig.~\ref{fig_training}.
In the receive processing, reference signals/pilots are used to train the NN of the RC for the learning-based symbol detection strategy while the remaining data symbols will go through the trained NN for symbol detection. 

\section{Introduced Approach}
\label{introduced_app}
By casting the symbol detection problem in a Bayesian framework, it can be formulated as the following maximum likelihood estimation problem,
\begin{align}
    \arg \max_{\boldsymbol {\tilde X}} P({\boldsymbol {\tilde X}}|{\boldsymbol {\tilde Y}}).
\end{align}
where $P(\cdot|\cdot)$ represents the joint posterior distribution. The joint posterior distribution can then be approximated via the naive Bayesian principle, i.e., 
\begin{align}
P({\boldsymbol {\tilde X}}|{\boldsymbol {\tilde Y}}) \sim \prod_{n_t\in 2N_t, n_c\in N_c} P_{n_t,n_c}({\tilde x}_{n_t,n_c}| {\boldsymbol {\tilde Y}})
\end{align}
where 
$P_{n_t,n_c}(\cdot|{\boldsymbol {\tilde Y}})$ stands for the marginal distribution of the $(n_t, n_c)^{\text{th}}$ entry of $\boldsymbol {\tilde X}$. Accordingly, we train NNs to approximate  $P_{n_t,n_c}(\cdot|{\boldsymbol {\tilde Y}})$, i.e.,
\begin{align}
f_{n_t,n_c}({\tilde x}_{n_t,n_c}; {\boldsymbol {\tilde Y}}) = P({\tilde x}_{n_t,n_c}|{\boldsymbol {\tilde Y}})
\end{align}
where $f_{n_t,n_c}({\tilde x}_{n_t,n_c}; {\boldsymbol {\tilde Y}})$ denotes the NN with $\tilde x \in \{-3,-1,+1,+3\}$ as the output and $\boldsymbol {\tilde Y}$ as the input. 
In the final step, the transmitted symbols can be estimated by choosing the maximum a posterior estimation, 
\begin{align}
\arg \max_{{\tilde x}_{n_t,n_c}} f_{n_t,n_c}({\tilde x}_{n_t,n_c}; {\boldsymbol {\tilde Y}}).
\end{align}
Therefore, the symbol estimation task is completed by collecting the NN outputs corresponding to all the entries of $\boldsymbol {\tilde X}$.

\subsection{Detection on a Binary Set}
\begin{figure}[h]
    \centering
    \includegraphics[width =1 \linewidth]{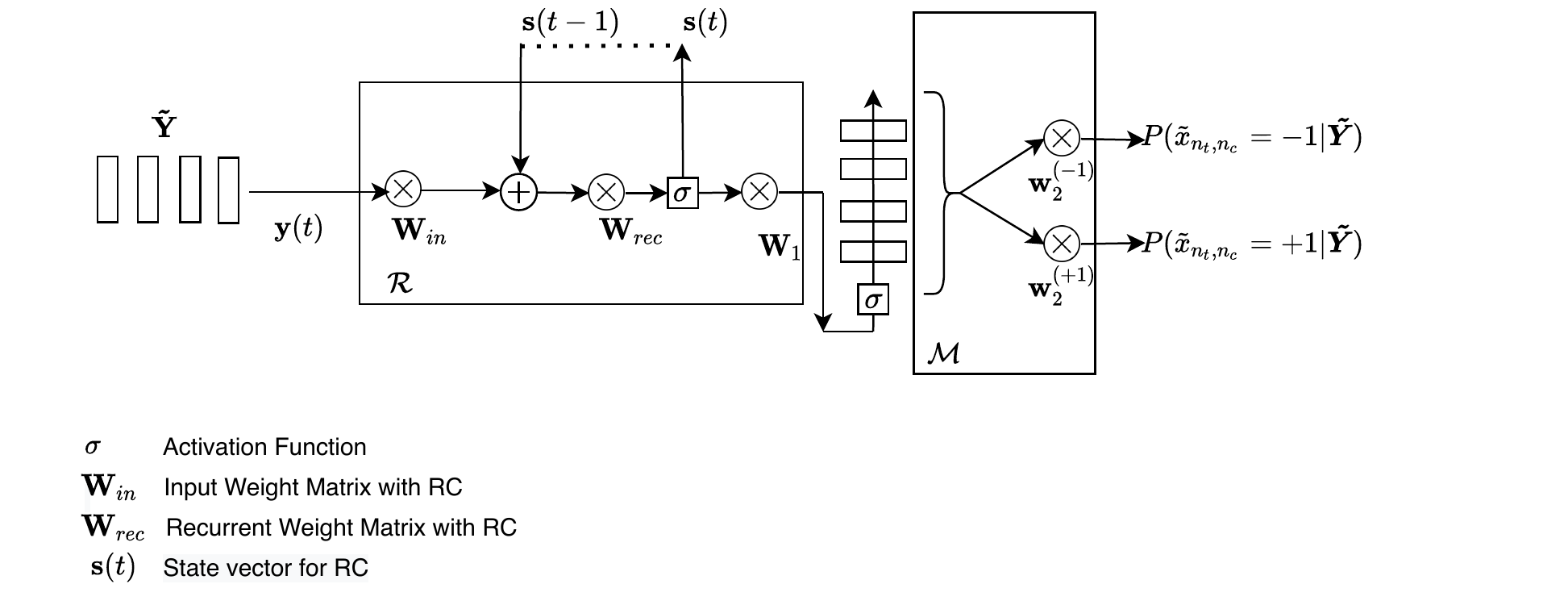}
    \caption{Binary detector $f_{n_t,n_c}({\tilde x}_{n_t, n_c}; {\boldsymbol {\tilde Y}})$}
    \label{fig_rc}
\end{figure}
For simplicity, we first consider the training of a binary decision NN for ${\tilde x}_{n_t,n_c}$, i.e., the NN only has two outputs. 
The underlying binary detector can be seen more clearly in Fig.~\ref{fig_rc}.
To be specific, the output ratio of this NN is an estimation of the marginal likelihood ratio of
\begin{align*}
{{P({\tilde x}_{n_t,n_c} = +1 |{\boldsymbol {\tilde Y}})}\over
{P({\tilde x}_{n_t,n_c} = -1 |{\boldsymbol {\tilde Y}})}} =
{f_{n_t,n_c}({\tilde x}_{n_t,n_c} = +1; {\boldsymbol {\tilde Y}})\over
f_{n_t,n_c}({\tilde x}_{n_t,n_c} = -1; {\boldsymbol {\tilde Y}})}
\end{align*}
For ease of discussion, we introduce ${\mathcal L}_{+-}$ as a notation of the likelihood ratio, i.e., 
\begin{align}
{\mathcal L}_{+-}(\boldsymbol {\tilde Y}) := {f_{n_t,n_c}({\tilde x}_{n_t,n_c} = +1; {\boldsymbol {\tilde Y}})\over
f_{n_t,n_c}({\tilde x}_{n_t,n_c} = -1; {\boldsymbol {\tilde Y}})}
\end{align}
To capture the time domain features of MIMO-OFDM signals, we design the NN as
\begin{align}
    \label{binary_nn}
    f_{n_t,n_c}({\tilde x}_{n_t,n_c}; {\boldsymbol {\tilde Y}}) := {\mathcal M}_{{\boldsymbol W}_2}({\mathcal R}_{{\boldsymbol W}_1}({\boldsymbol {\tilde Y}}))
\end{align}
where $\mathcal M$ is a single-layer perceptron (feedforward NN) with a soft-max output, the coefficients to learn are denoted as ${\boldsymbol W}_2 := [{\boldsymbol w}_2^{(-1)}, {\boldsymbol w}_2^{(+1)}]$, in which ${\boldsymbol w}_2^{(+1)}$, ${\boldsymbol w}_2^{(-1)}$ are the coefficients corresponding to the positive output and the negative output respectively. $\mathcal R$ represents a RNN. We fix the weights of the input layer and state transition layer of $\mathcal R$, and only train the output layer ${\boldsymbol W}_1$. This particular RNN structure follows the framework of reservoir computing (RC). The input and output dimension of the RNN is the exactly $N_r$. In general, RC is characterized by a \textbf{state equation} and an \textbf{output equation} \cite{lukovsevivcius2009reservoir}. 
In this paper, we choose the state equation as a first order Markov process, where the state are involved by input stimulus and initial states. The output equation is a linear mapping which readouts the internal states. Furthermore, we add a hyperbolic tangent (Tanh) function as the activation function between $\mathcal R$ and $\mathcal M$.



\subsection{Detection on a Non-Binary Set}

\begin{figure}
    \centering
    \includegraphics[width =1 \linewidth]{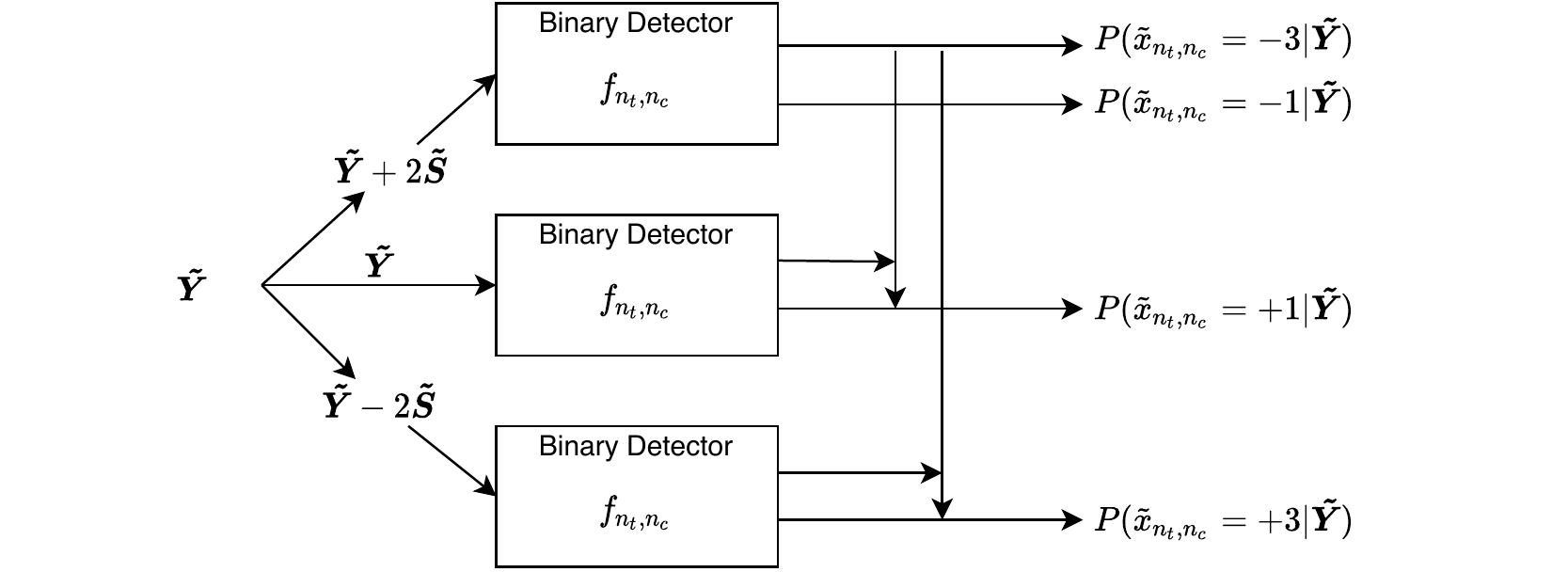}
    \caption{Multiple-class detector generated by binary detector.}
    \label{mutiple_class_nn}
\end{figure}
To conduct symbol detection on a non-binary set (e.g., $\{-3, -1, +1, +3\}$), we utilize the following shifting process to construct element-wise likelihood ratios introduced in \cite{zhou2020learning},
\begin{align*}
    {{P({\tilde x}_{n_t,n_c} = -1 |{\boldsymbol {\tilde Y}})}\over
    {P({\tilde x}_{n_t,n_c} = -3 |{\boldsymbol {\tilde Y}})}} =
    {\mathcal L}_{+-}(\boldsymbol {\tilde Y} + 2{\boldsymbol {\tilde S}}_{n_t,n_c})\\
    {{P({\tilde x}_{n_t,n_c} = +3 |{\boldsymbol {\tilde Y}})}\over
    {P({\tilde x}_{n_t,n_c} = +1 |{\boldsymbol {\tilde Y}})}} =
    {\mathcal L}_{+-}(\boldsymbol {\tilde Y} - 2{\boldsymbol {\tilde S}}_{n_t,n_c}) 
\end{align*}
where  
\begin{align*}
{\boldsymbol {\tilde S}}_{n_t,n_c} := \begin{bmatrix}
{\text{Re}}({\boldsymbol {S}}_{n_t,n_c})\\
{\text{Im}}({\boldsymbol {S}}_{n_t,n_c})
\end{bmatrix}
\end{align*}
\begin{equation*}
{\boldsymbol S}_{n_t,n_c} := \left\{
\begin{aligned}
\sum_{\ell = 0}^{L-1} {\boldsymbol H}_{n_t,:}[\ell] ({{\boldsymbol F} {\boldsymbol J}_{\ell}})_{:,n_c}, {\text {if }} {\tilde x}_{n_t,n_c} = {\text{Re}}(x_{n_t,n_c}) \\
\sum_{\ell = 0}^{L-1} j{\boldsymbol H}_{n_t,:}[\ell] ({{\boldsymbol F} {\boldsymbol J}_{\ell}})_{:,n_c}, {\text {if }} {\tilde x}_{n_t,n_c} = {\text{Im}}(x_{n_t,n_c}) 
\end{aligned}
\right.
\end{equation*}
Note that the received signal $\boldsymbol {\tilde Y}$ corresponds to transmitted symbols modulated using the general QAM constellation. Therefore, the posterior estimation can be obtained by solving the equation below:
\begin{equation*}
\begin{aligned}
    &\sum_{a = \{-3,-1,+1,+3\}} P({\tilde x}_{n_t,n_c} = a| {\boldsymbol {\tilde Y}}) = 1\\
    &P({\tilde x}_{n_t,n_c} = -1| {\boldsymbol {\tilde Y}}) = P({\tilde x}_{n_t,n_c} = -3| {\boldsymbol {\tilde Y}}){\mathcal L}_{+-}(\boldsymbol {\tilde Y} + 2{\boldsymbol S}_{ij})\\
    &P({\tilde x}_{n_t,n_c} = +1| {\boldsymbol {\tilde Y}}) = P({\tilde x}_{n_t,n_c} = -1| {\boldsymbol {\tilde Y}}){\mathcal L}_{+-}(\boldsymbol {\tilde Y})\\
    &P({\tilde x}_{n_t,n_c} = +3| {\boldsymbol {\tilde Y}}) = P({\tilde x}_{n_t,n_c} = +1| {\boldsymbol {\tilde Y}}){\mathcal L}_{+-}(\boldsymbol {\tilde Y} +-2{\boldsymbol S}_{ij})
\end{aligned}
\end{equation*}
Therefore, the complete NN for multi-class detection can be constructed as in Fig. \ref{mutiple_class_nn}.

\subsection{Training Set}
As outlined in the previous discussion, the symbol detection NN for a general QAM constellation can be constructed by shifting individual binary detection NNs. 
Therefore, training the NN in Fig. \ref{mutiple_class_nn} is equivalent to training the binary decision neural network (\ref{binary_nn}). 
Assume we have $Q$ training samples,
\begin{align}
\label{time_training}
\left\{({\boldsymbol {\tilde X}}_1, {\boldsymbol {\tilde Y}}_1), ({\boldsymbol {\tilde X}}_2, {\boldsymbol {\tilde Y}}_2),
\cdots, 
({\boldsymbol {\tilde X}}_Q, {\boldsymbol {\tilde Y}}_Q)\right\}
\end{align}
where entries of ${\boldsymbol {\tilde X}}_q$ come from a general QAM constellation, the corresponding binary training sets for $(n_t,n_c)$ can be created according to:
\begin{equation}
\label{binary_trainset}
\begin{aligned}
    \left(+1, {\boldsymbol {\tilde Y}}_k-{{\tilde x}_{n_t,n_c}} {\boldsymbol {\tilde S}}_{n_t,n_c} + {\boldsymbol {\tilde S}}_{n_t,n_c}\right)\\
    \left(-1, {\boldsymbol {\tilde Y}}_k-{{\tilde x}_{n_t,n_c}} {\boldsymbol {\tilde S}}_{n_t,n_c} - {\boldsymbol {\tilde S}}_{n_t,n_c}\right) 
\end{aligned}
\end{equation}
through $({\boldsymbol {\tilde X}}_k, {\boldsymbol {\tilde Y}}_k)$. For wireless systems where the training set is limited, we set $Q = N_t$ to be compatible with the standard orthogonal pilot structures adopted in modern cellular standards such as 3GPP LTE-Advanced and 5G.

\subsection{Learning Tricks}

To mitigate the issues of over-fitting and slow training convergence, the following specific techniques are employed in our experiments:
\begin{itemize}
\item The weight ${\boldsymbol W}_1$ is shared across all sub-carriers in the frequency domain. 
\item The weight ${\boldsymbol W}_2$ is initialized according to ${\boldsymbol w}_2^{(+1)} = {\boldsymbol w}_2^{(-1)} = ({\boldsymbol I}_{N_r}\otimes{\boldsymbol F})_{i+jN, :}$. This particular initialization essentially follows from the Fourier transformation embedded in the generation of OFDM signals.
\item The binary detector $f_{n_t, n_c}$ can be trained by utilizing the training sets from the neighboring sub-carriers. This exploits the channel correlation between neighboring sub-carriers in frequency-selective fading channels. 
\end{itemize}

\section{Performance Evaluation}
\label{exp}
In this section, we provide performance evaluations of the introduced symbol detection approach. 
The modulation scheme used to generate $\boldsymbol X$ is $16$-QAM.
The underlying MIMO channel is assumed to be block fading. 
To be specific, the channel coefficients of the MIMO-OFDM system are generated using parameter configurations in the urban outdoor-to-indoor NLOS scenario of the WINNER II channel model~\cite{meinila2009winner} and are assumed to stay invariant in a block of $Q + N_d$ OFDM symbols. 
Uniform Linear Array (ULA) antennas are used at the transmitter and receiver with a half-wavelength spacing. 
The simulation parameters in the performance evaluation are set as follows: $L=12$, $N_r = 2$, $N_t = 2$, $N_{c} = 256$, $N_{cp} = 25$, $Q = 2$, and $N_d = 98$.
The number of neurons in the RC is set to $128$ with an input buffer of length $32$ as the windowed-ESN structure discussed in~\cite{zhou2019}.

\begin{figure}
    \centering
    \includegraphics[width = 1\linewidth]{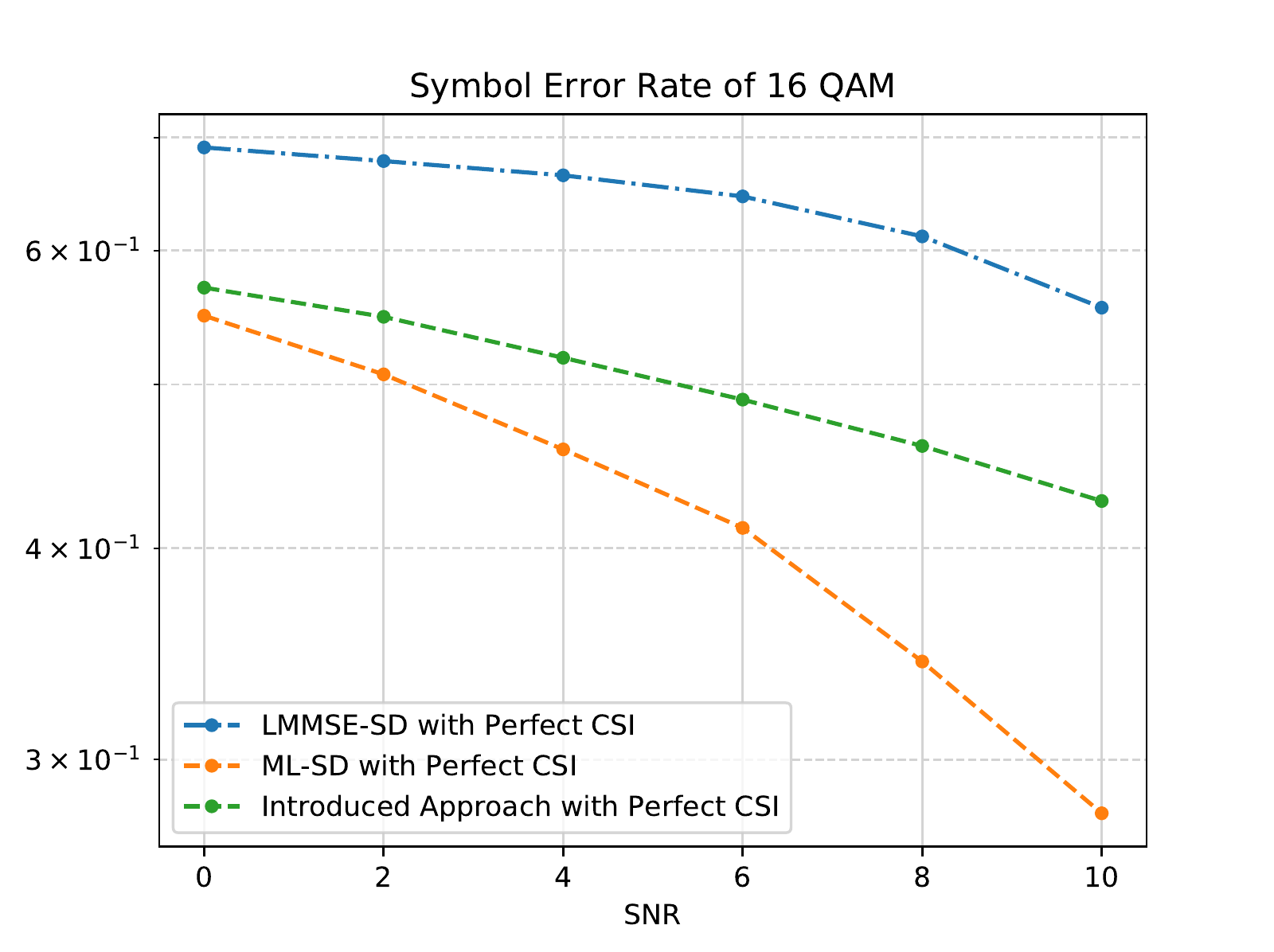}
    \caption{Symbol error rate comparison under perfect CSI. }
    \label{perfect_CSI}
\end{figure}

\begin{figure}
    \centering
    \includegraphics[width = 1\linewidth]{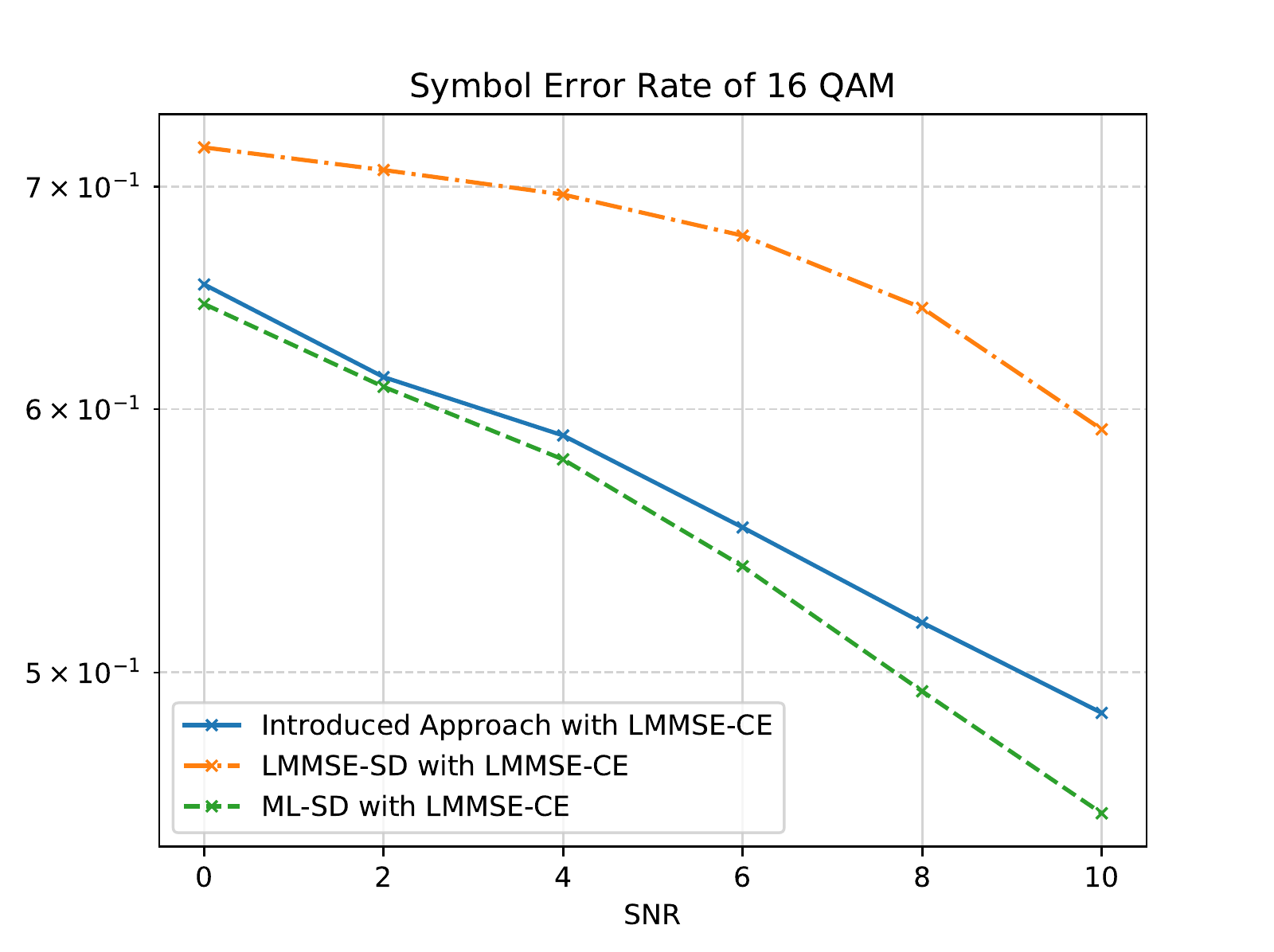}
    \caption{Symbol error rate comparison under LMMSE-based estimated CSI. }
    \label{imperfect_CSI}
\end{figure}

To evaluate the symbol detection performance, we first compare the introduced approach to conventional methods assuming perfect CSI. This result is illustrated in Fig. \ref{perfect_CSI}. As we can observe, the introduced method outperforms the conventional linear minimum mean square estimation (LMMSE) strategy while is strictly inferior to the optimal maximum likelihood (ML) detection strategy. In a more relevant scenario, we conduct the CSI estimation first using the available reference signals/pilots.
The underlying performance evaluation is shown in Fig. \ref{imperfect_CSI} where we conduct LMMSE-based channel estimation to obtain the estimated CSI. 
The results suggest that the introduced method can offer detection performance that is close to the optimum ML method especially in the low SNR regime.
Note that the complexity of RC-based strategy is even lower than that of LMMSE-based strategy~\cite{RC_Complexity}.


\begin{figure}
    \centering
    \includegraphics[width = 1\linewidth]{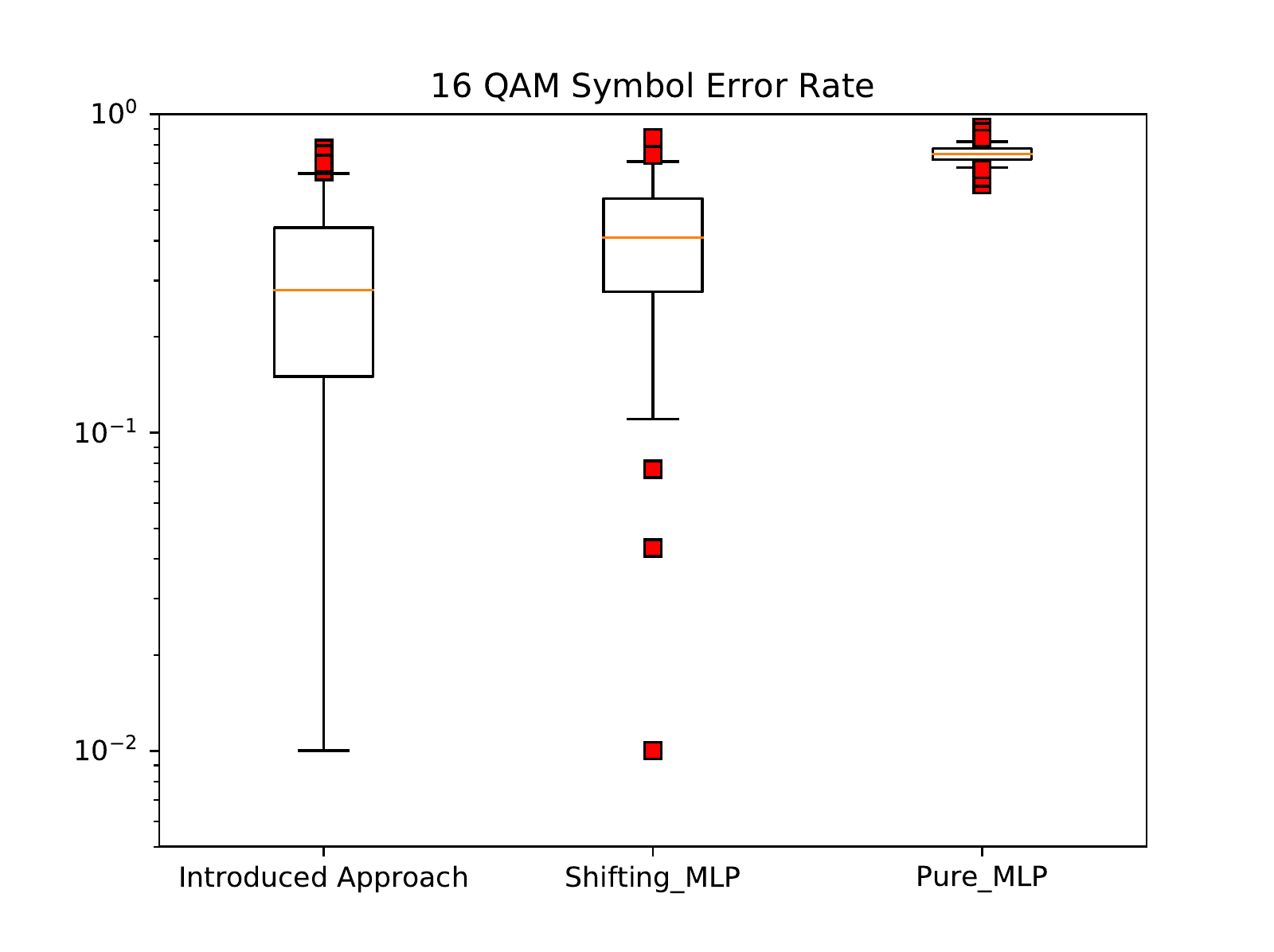}
    \caption{SNR = 10 dB, Training Set Size = $2$ MIMO-OFDM Symbols}
    \label{BER_Box_plot}
\end{figure}

Finally, we present the performance comparison under alternate NN architectures. 
This comparison uses the metric of generalization error which is characterized by the symbol error rate (SER) in this case, as demonstrated in Fig. \ref{BER_Box_plot}. 
The label `shifting MLP' represents a structure similar to Fig.~\ref{binary_nn}, but using a multilayer perceptron (MLP) in the time domain instead of the RC. 
The training scheme is the same as in the introduced network. 
The label `pure MLP' represents an MLP that performs the symbol detection directly instead of employing the knowledge of structure to build the multi-class symbol detector. 
The results demonstrate that our method which includes a strong inductive bias can achieve a better generalization performance compared to other generic non structure-aware NN architectures illustrating the power of RC and structure knowledge in wireless systems.

\section{Conclusion}
\label{conclusion}
A NN-based symbol detector for MIMO-OFDM systems is introduced in this paper. 
The NN incorporates structural knowledge of MIMO-OFDM signals that characterizes features in time, frequency and constellation where the training overhead is significantly reduced due to the inductive bias. 
The time domain information is captured through the RC while the frequency domain information is reflected in the initialization of RC's output layer. 
The structure of the modulation scheme is incorporated by a shifting process which generalizes binary symbol detectors to multi-class detectors. 
The training of the associated network weights is conducted through backpropagation. 
Experimental results show that the introduced NN-based detection can perform close to ML detection under estimated CSI.
Since the complexity of the introduced RC-based strategy is relatively low, it provides a promising symbol detection strategy for MIMO-OFDM systems.


\section*{Appendix}
\subsection{Equivalent Real-value Signal Model}
Let $g(\cdot)$ represent the entire linear mapping in (\ref{MIMO_OFDM0}), 
\begin{align}
    {\boldsymbol {\tilde Y}} = g({\boldsymbol {\tilde X}}) + {\boldsymbol {\tilde N}}.
\end{align}
The real-valued form of the above observation equation can be written as the following composite function
\begin{align*}
\label{eq_observation}
\begin{bmatrix}
{\text{Re}}(\boldsymbol {Y})\\
{\text{Im}}(\boldsymbol {Y})
\end{bmatrix}
= 
\sum_{\ell=0}^{L-1} g^{(\ell)} \left (f^{(\ell)} \left (\begin{bmatrix}
{\text{Re}}(\boldsymbol {X})\\
{\text{Im}}(\boldsymbol {X})
\end{bmatrix}\right ) \right ) + 
\begin{bmatrix}
{\text{Re}}(\boldsymbol {N})\\
{\text{Im}}(\boldsymbol {N})
\end{bmatrix}
\end{align*}
where $f^{(\ell)}(\cdot)$ is defined as
\begin{align*}
\begin{bmatrix}
{\text{Re}}({\boldsymbol {XFJ}_{\ell}})^T\\
{\text{Im}}({\boldsymbol {XFJ}_{\ell}})^T
 \end{bmatrix} 
= \begin{bmatrix}
{\text{Re}}({\boldsymbol {FJ}_{\ell}})^T,
-{\text{Im}}({\boldsymbol {FJ}}_{\ell})^T\\
{\text{Im}}({\boldsymbol {FJ}_{\ell}})^T,
{\text{Re}}({\boldsymbol {FJ}_{\ell}})^T
 \end{bmatrix}
 \begin{bmatrix}
{\text{Re}}({\boldsymbol {X}})^T\\
{\text{Im}}({\boldsymbol {X}})^T
 \end{bmatrix}
\end{align*}
and $g^{(\ell)}(\cdot)$ is defined as
\begin{align*}
&\begin{bmatrix}
{\text{Re}}(\boldsymbol {Y})\\
{\text{Im}}(\boldsymbol {Y})
\end{bmatrix}
 = 
\begin{bmatrix}
{\text{Re}}({\boldsymbol {H}}[\ell]), -{\text{Im}}({\boldsymbol {H}}[\ell])\\
{\text{Im}}({\boldsymbol {H}}[\ell]), {\text{Re}}({\boldsymbol {H}}[\ell])
 \end{bmatrix}\begin{bmatrix}
{\text{Re}}({\boldsymbol {XFJ}}_{\ell})\\
{\text{Im}}({\boldsymbol {XFJ}}_{\ell})
 \end{bmatrix}.\\
\end{align*}
\subsection{Channel Estimation}
Channel estimation is conducted in the frequency domain. 
Multiplying the discrete Fourier transform (DFT) matrix ${\boldsymbol F}^H$ on both sides of (\ref{MIMO_OFDM0}), we have 
\begin{align}
{\boldsymbol Y}{\boldsymbol F}^H = \sum_{\ell=0}^{L-1} \boldsymbol{H}[\ell] \boldsymbol{X} {\text{diag}}({\boldsymbol l}_{\ell})  +{\boldsymbol N}{\boldsymbol F}^H
\end{align}
where ${\boldsymbol l}_{\ell}$ is the vector of eigenvalues of ${\boldsymbol J}_{\ell}$, since the eigenvectors of a circulant matrix is the Fourier matrix, and the ${\text{diag}}()$ operator converts its vector argument to a diagonal matrix. Accordingly, the $j^{\text{th}}$ column of $\boldsymbol X$ can be written as
\begin{align}
\left({\boldsymbol Y}{\boldsymbol F}^H\right)_{:,j} = \sum_{\ell=0}^{L-1} l_{\ell}(j)\boldsymbol{H}[\ell] \boldsymbol{x}_j +({\boldsymbol N}{\boldsymbol F}^H)_{:,j}
\end{align}
where $\sum_{\ell=0}^{L-1} l_{\ell}(j)\boldsymbol{H}[\ell]$ stands for the equivalent channel on the $j^{\text{th}}$ sub-carrier which can be estimated via the LMMSE method using the $Q$ symbols from the training set in (\ref{time_training}).

\bibliographystyle{./bibliography/IEEEtran}
\bibliography{./bibliography/IEEEabrv,./bibliography/ref}

\end{document}